\documentclass[11pt]{cernrep}
\usepackage{psfig}
\usepackage{epsfig}
\usepackage{axodraw}
\usepackage{graphicx}
\usepackage{here}
\usepackage{graphics}
\usepackage{dcolumn}
\usepackage{amsmath}
\usepackage{rotating}
\usepackage{amssymb}
\usepackage{amsbsy}
\usepackage{mcite}

\newcommand{\tev}{{\rm Te}\kern-1.pt{\rm V}}
\newcommand{\gev}{{\rm Ge}\kern-1.pt{\rm V}}
\newcommand{\mev}{{\rm Me}\kern-1.pt{\rm V}}
\newcommand{\kev}{{\rm Ke}\kern-1.pt{\rm V}}
\newcommand{\gevsq}{\mbox{$\mathrm{{\rm Ge}\kern-1.pt{\rm V}}^2$}}
\newcommand{\gevmsq}{\mbox{$\mathrm{{\rm Ge}\kern-1.pt{\rm V}}^{-2}$}}

\newcommand{\mayor} {\mbox{\raisebox{-0.4ex}
{$\;\stackrel{>}{\scriptstyle \sim}\;$}}}

\newcommand{\sla}[1]{/\!\!\!#1}
\begin{document}

\pagestyle{myheadings}

{
\noindent
{\Large \bf Prospects for Higgs Searches via VBF at the LHC with the ATLAS Detector} \\[0.5cm]
{\it K.~Cranmer, Y.Q.~Fang, B.~Mellado, S.~Paganis, W.~Quayle and
Sau~Lan~Wu}

\begin{abstract}
We report on the potential for the discovery of a Standard Model
Higgs boson with the vector boson fusion mechanism in the mass
range $115<M_H<500\,\gev$ with the ATLAS experiment at the LHC.
Feasibility studies at hadron level followed by a fast detector
simulation have been performed for $H\rightarrow
W^{(*)}W^{(*)}\rightarrow l^+l^-\sla{p_T}$,
$H\rightarrow\gamma\gamma$ and $H\rightarrow ZZ\rightarrow
l^+l^-q\overline{q}$. The results obtained show a large discovery
potential in the range $115<M_H<300\,\gev$. Results obtained with
multivariate techniques are reported for a number of channels.
\end{abstract}

\section{Introduction}
\label{sec:introduction}

In the Standard Model (SM) of electro-weak and strong
interactions, there are four types of  gauge vector bosons (gluon,
photon, W and Z) and twelve types of fermions (six quarks and six
leptons)~\cite{np_22_579,prl_19_1264,sal_1968_bis,pr_2_1285}.
These particles have been observed experimentally. The SM also
predicts the existence of one scalar boson, the Higgs
boson~\cite{pl_12_132,prl_13_508,pr_145_1156,prl_13_321,prl_13_585,pr_155_1554}.
The observation of the Higgs boson remains one of the major
cornerstones  of the SM. This is a primary focus of the ATLAS
Collaboration~\cite{LHCC99-14}.

The  Higgs at the LHC is produced predominantly via gluon-gluon
fusion~\cite{prl_40_11_692}. For Higgs masses, $M_H$,  such that
$M_H>100\,\gev$, the second dominant process is vector boson
fusion (VBF)~\cite{pl_136_196,pl_148_367}.

Early  analyses performed at the parton level with the decays
$H\rightarrow W^{(*)}W^{(*)}$, $\tau^+\tau^-$  and $\gamma\gamma$
via VBF indicated that this mechanism could produce strong
discovery modes in the range of the Higgs mass
$115<M_H<200\,\gev$~\cite{pr_160_113004,pl_503_113,pr_61_093005,JHEP_9712_005}.
The ATLAS collaboration has performed feasibility studies for
these decay modes including more detailed detector description and
the implementation of initial-state and final-state parton showers
(IFSR), hadronization and multiple
interactions~\cite{SN-ATLAS-2003-024}.

Here, we present an update of the potential of observing the Higgs
boson via VBF with $H\rightarrow W^{(*)}W^{(*)}\rightarrow
l^+l^-\sla{p_T}$, where $\sla{p_T}$ stands for missing transverse
momentum carried by neutrinos, reported
in~\cite{SN-ATLAS-2003-024}.  This analysis has been extended to
larger Higgs masses. Also, we investigated the prospects of
observing a SM Higgs boson with $H\rightarrow\gamma\gamma$ and
$H\rightarrow ZZ\rightarrow l^+l^-q\overline{q}$. Results obtained
with multivariate techniques are reported for a number of
channels. Finally, the status of the overall SM Higgs discovery
potential of the ATLAS detector is presented.

\section{Experimental Signatures}
\label{sec:expsig}

The VBF mechanism displays a number of distinct features, which
may be exploited experimentally to suppress SM backgrounds: Higgs
decay products are accompanied by two energetic forward jets
originating from incoming quarks and suppressed jet production in
the central region is expected due to the lack of color flow
between the initial state quarks. In this paper, tagging jets are
defined as the highest and next highest transverse momentum,
$P_T$, jets in the event. A number of variables were used in the
VBF analyses reported in this paper: $P_T$ of the leading and the
sub-leading jets, $P_{Tj_1}$ and $P_{Tj_2}$, pseudorapidity,
$\eta$, of the leading and sub-leading jets, $\eta_{j_1}$ and
$\eta_{j_1}$, with
$\Delta\eta_{j_1j_2}=\left|\eta_{j_1}-\eta_{j_2}\right|$, the
difference of tagging jets' azimuthal angles,
$\Delta\phi_{j_1j_2}$ and tagging jets' invariant mass,
$M_{j_1j_2}$. The tagging jets are required to be in opposite
hemispheres ($\eta_{j_1}\eta_{j_2}<0$).

In Sections~\ref{sec:hww} and~\ref{sec:hzz} a number of variables
are used: pseudorapidity and azimuthal angle difference between
leptons, $\eta_{ll}$ and $\phi_{ll}$, and di-lepton invariant
mass, $M_{ll}$. In Section~\ref{sec:hgg} the following variables
are used: $P_T$ of the leading and sub-leading $\gamma$'s,
$P_{T\gamma_1}$ and $P_{T\gamma_2}$, pseudorapidity and azimuthal
angle difference between $\gamma$'s, $\eta_{\gamma\gamma}$ and
$\phi_{\gamma\gamma}$, and di-$\gamma$ invariant mass,
$M_{\gamma\gamma}$.

The following decay chains have been considered in the analysis:
$H\rightarrow W^{(*)}W^{(*)}\rightarrow l^+l^-\sla{p_T}$, $H
\rightarrow \gamma \gamma$ and $H\rightarrow ZZ\rightarrow
l^+l^-q\overline{q}$. A number of relevant experimental aspects
have been addressed in detail
in~\cite{LHCC99-14,SN-ATLAS-2003-024} and will not be touched upon
in this work: triggering, lepton and photon identification, fake
lepton and photon rejection, jet tagging, central jet veto and
b-jet veto efficiencies.\footnote{The central jet and b-jet vetoes
are applied if a jet (b-jet) with $P_T>20\,\gev$ is found in the
range $\left|\eta\right|<3.2$ and $\left|\eta\right|<2.5$,
respectively.}

\section{Signal}
\label{sec:signal}

 The Born level cross-sections for the VBF process have been
calculated using the PYTHIA
package~\cite{cpc_82_74,cpc_135_238}.\footnote{The results from
PYTHIA agree with the calculation provided by VV2H~\cite{VV2H} by
better than $3\,\%$.} The results are given in
Tables~\ref{t:sig_br}-\ref{t:sig_br2} for different Higgs masses.
The signal (and background) Born level cross-sections have been
computed using the CTEQ5L structure function
parametrization~\cite{epj_12_375}. The products of the signal
cross-sections and the branching ratios of the Higgs boson into
$W^{(*)}W^{(*)}$, $\gamma\gamma$, and $ZZ$, which have been
calculated using the programme HDECAY~\cite{cpc_108_56}, are also
included in Table~\ref{t:sig_br}-\ref{t:sig_br2}.

\begin{table}[t]
\begin{center}
\footnotesize
\begin{tabular}{l r || c  c  c  c  c  c  c  }
\hline \hline
$m_H$ &  $(\gev)$  & 120 & 130 & 140 & 150 & 160 & 170 & 180\\
\hline \hline
$\sigma (q\overline{q} H)$ & (pb) & 4.29 & 3.97 & 3.69 & 3.45 & 3.19 & 2.95 & 2.80 \\
\hline $\sigma \cdot BR (H \rightarrow W^{(*)}W^{(*)})$ & (fb) &
522 & 1107 & 1771 & 2363 & 2887 & 2850 & 2618 \\
$\sigma \cdot BR (H \rightarrow \gamma \gamma)$ & (fb) &
 9.38 & 8.89 & 7.14 & 4.76 & 1.71 & -  & -   \\
\hline \hline
\end{tabular}
\footnotesize \caption{\small \it Total vector boson fusion
production cross-sections, $\sigma (q\overline{q}H)$,  $\sigma
\cdot BR (H \rightarrow W^{(*)}W^{(*)})$ and $\sigma \cdot BR (H
\rightarrow \gamma \gamma )$ for a low mass Higgs. The
cross-sections have been computed using the CTEQ5L structure
function parametrization. }\label{t:sig_br}
\end{center}
\end{table}

\begin{table}[t]
\begin{center}
\footnotesize
\begin{tabular}{l r || c  c  c  c  c  c  c  c }
\hline \hline
$m_H$ &  $(\gev)$ & 190 & 200 & 250 & 300 & 350 & 400 & 450 & 500 \\
\hline \hline
$\sigma (q\overline{q} H)$ & (pb) & 2.58 & 2.44 & 1.82 & 1.38 & 1.06 & 0.83 & 0.66 & 0.53 \\
\hline $\sigma \cdot BR (H \rightarrow WW)$ & (fb) &
2005 & 1793 & 1276 & 954 & 721 & 488 & 363 & 289 \\
$\sigma \cdot BR (H \rightarrow ZZ)$ & (fb) &
 562 & 637 & 542 & 424 & 332 & 227 & 172 & 138 \\
\hline \hline
\end{tabular}
\footnotesize \caption{\small \it Total vector boson fusion
production cross-sections, $\sigma \cdot BR (H \rightarrow WW)$
and $\sigma \cdot BR (H \rightarrow ZZ )$ for a heavier Higgs. The
cross-sections have been computed using the CTEQ5L structure
function parametrization. }\label{t:sig_br2}
\end{center}
\end{table}

The impact of initial and final state QCD radiation,
hadronization, multiple interactions and underlying event were
simulated with PYTHIA6.1~\cite{cpc_82_74,cpc_135_238}. The signal
and background simulation used the package ATLFAST~\cite{ATLFAST}
in order to simulate the response of the ATLAS detector.

\section{The $H\rightarrow W^{(*)}W^{(*)}\rightarrow l^+l^-\sla{p_T}$ Mode}
\label{sec:hww}

A study of this mode at hadron level followed by a fast simulation
of the ATLAS detector was first performed
in~\cite{ButtarHarperJakobs}. In this Section we report on a
re-analysis over a broader mass range $115<M_H<500\,\gev$.
Additionally, the treatment of the main background process is
improved in the present analysis.

\subsection{Background Generation}
\label{sec:wwbackround}

\subsubsection{$t\overline{t}$ Production Associated with Jets}\
\label{sec:tt} The production of $t\overline{t}$ associated with
one jet, $t\overline{t}j$, was identified as the main background
process for this mode~\cite{pr_160_113004,pl_503_113}. Early
parton level analyses were based on $t\overline{t}j$ Leading Order
(LO) Matrix Element (ME) calculation. In order to assess
hadronization and detector effects, it is necessary to interface
the fixed order ME calculations with a parton shower in a
consistent way. Here we use a Next-to-Leading-Order description of
the $t\overline{t}$ ME matched with parton shower provided within
the MC@NLO package, which avoids double-counting and allows for a
smooth matching between hard and soft/collinear emission
regions~\cite{JHEP_0206_029,JHEP_0308_007}.
\begin{figure}[t]
    \begin{center}
        \vspace{-0.5cm}
      \mbox{\psfig{figure=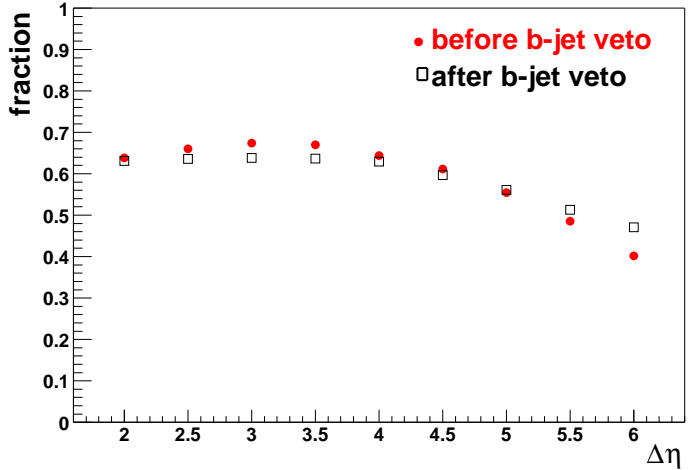,width=4.0in}}
\vspace{-0.2cm}
      \caption{
Fraction of events for which either the leading or the sub-leading
jet is a b-jet as a function of the cut on $\Delta\eta_{j_1j_2}$
before and after the application of a b-jet veto.}
        \label{1B}
    \end{center}
\end{figure}
In MC@NLO hard emissions are treated as in NLO calculations while
soft/collinear emissions are handled by the MC simulation
(HERWIG6.5 in this case) with the MC logarithmic accuracy: the
$t\overline{t}$ rates are known to NLO while the parton shower
part preserves unitarity. Comparisons between MC@NLO and LO event
generators PYTHIA6.2~\cite{cpc_82_74,cpc_135_238} and
HERWIG6.5~\cite{JHEP_0101_010,HERWIG6.5} show that, within the
MC@NLO approach, the low $P_T$ region is dominated by the parton
shower prescription, while at higher $P_T$ the NLO calculation
dominates predicting a significantly higher $P_T$ for the
$t\overline{t}$ system.

PYTHIA6.2 predicts a softer $P_T$ distribution with strong
differences in the high $P_T$ region ($P_T>100\,\gev$) with
respect to the NLO prediction. It was also found that all three
models give similar b-jet $P_T$ distribution.

The MC@NLO  description of the second jet from the
$t\overline{t}jj$ process was tested against a LO
$t\overline{t}jj$ ME calculation using
MadGraphII~\cite{pc_81_357,hep-ph_0208156} interfaced to
HERWIG6.5~\cite{Wisc_soft}. To reduce the double-counting in the
HERWIG6.5 interface with MadGraphII, the parton shower cutoff was
set to the $P_T$ of the lowest $P_T$ QCD parton in the ME
calculation. The resulting $P_T$ distribution comparison showed
that MC@NLO predicts a sub-leading non-b jet which is in good
agreement for $P_T>50\,\gev$ with the MadGraphII $t\overline{t}jj$
ME calculation. In conclusion, MC@NLO also provides a
reasonable description of the sub-leading radiation.

MC@NLO was used to define a  $t\overline{t}j$ control sample via
an event selection similar to the one used
in~\cite{pr_160_113004,pl_503_113,pr_61_093005,JHEP_9712_005}.
The dependence of various kinematic distributions on
$\Delta\eta_{j_1j_2}$ was evaluated. In a large fraction
($\simeq20\%$) of events with small values of
$\Delta\eta_{j_1j_2}$, both leading jets are b-jets. For
$\Delta\eta_{j_1j_2}>3.5$ about $65\%$ of the events have just one
of the two leading jets being a b-jet (see Figure~\ref{1B}). This fraction is clearly
dominated by $t\overline{t}j$ where the extra jet is hard. The
rest of the events were examined and about $30\%$ were found to have two
leading jets that are non-b-jets. These events are dominated by
$t\overline{t}jj$ where the two radiated partons are hard.

The results presented here show a small dependence of the jet
topology on the b-jet veto.  Only the third most energetic jet is
affected but the reduction of the fraction of events for which the
third jet is a b-jet is nearly constant as a function of the cut
on $\Delta\eta_{j_1j_2}$. According to these results, it is
possible to define a control sample in the early stages of data
taking with ATLAS to study properties of the $t\overline{t}$
process (for instance, normalization, central jet veto, b-jet
veto). One would like to use the part of the phase space which is
dominated by $t\overline{t}j$ and this is clearly the region for
which the separation of the tagging jets is
$\Delta\eta_{j_1j_2}\mayor3.5$. For a $<10\%$ systematic error in
the normalization of the $t\overline{t}j$ background about
$300-500$ pb$^{-1}$ of integrated luminosity will be
needed.\footnote{More details on this work are available
in~\cite{ATL-COM-PHYS-2003-043}.}

\subsubsection{Other Background Processes}
Other background processes were
considered~\cite{SN-ATLAS-2003-024}:
\begin{itemize}
\item Electro-weak $WWjj$ production; a quark scattering process
mediated by a vector boson, where the W bosons are produced and
decay leptonically.  This process is the second-dominant
background for most masses. To model this process, we use a
ME~\cite{zeppenfeld_1} that has been interfaced to
PYTHIA6.1~\cite{Mazini}. \item QCD $WWjj$ production. For this
process, we use the generator provided in PYTHIA6.1. \item
Electro-weak $Zjj$ production.  A $Z$ boson is produced in a
weak-boson-mediated quark-scattering process and decays into
$\tau$'s, which in turn decay leptonically.  This process was
modelled using a LO ME from the MadCUP project~\cite{MadCUP}.
\item QCD $Zjj$ production.  For this process, we use a LO ME from
the MadCUP project. As before, we consider events where
$Z\rightarrow\tau^+\tau^-$, $\tau\rightarrow l\nu\nu$. \item QCD
$Zjj$ production with $Z\rightarrow l^+l^-$ and $l=e,\mu$. This
background can be reduced substantially by requiring a minimum
missing $P_{T}$. However, it cannot be ignored because of its
large cross-section. We model this process with the generator
provided within PYTHIA6.1.\footnote{In the final version of this
work this process will be treated with a LO ME provided within
MadCUP.}
\end{itemize}

\subsection{Event Selection}
\label{sec:hwwevsel}

\begin{table}[t]
\begin{center}
\begin{tabular}{l || c c c c c c}
\hline \hline
Cut               & VBF       & $t\overline{t}$       & EW $WW$ & QCD $WW$        & EW $Zjj$        & QCD $Zjj$ \\
\hline \hline
{\bf a}     & 33.2  & 3.34$\times 10^3$      & 18.2  & 670   & 11.6  & 2.15$\times 10^3$     \\
{\bf b}     & 13.1  & 128   & 11.1  & 3.58  & 3.19  & 66.9  \\
{\bf c}     & 12.4  & 117   & 10.5  & 3.31  & 1.13  & 19.6  \\
{\bf d}     & 10.1  & 85.1  & 7.74  & 0.95 & 0.96  & 8.55  \\
{\bf e}     & 7.59  & 13    & 5.78  & 0.69 & 0.90 & 6.01  \\
{\bf f}     & 5.67  & 2.26  & 1.03  & 0.16 & 0.27 & 0.92 \\
{\bf g}     & 4.62  & 1.12  & 0.44 & 0.1        & 0.01        & 0.02        \\
\hline \hline
\end{tabular}
\caption{Cut flow for $M_{H}=160\,\gev$ in the $e-\mu$ channel.
Effective cross-sections are given in fb.  The event selection
presented in Section~\ref{sec:hwwevsel} is used. MC@NLO was used
to estimate the contribution from $t\overline{t}$ production (see
Section~\ref{sec:tt}) }\label{tab:mcanloCuts}
\end{center}
\end{table}

Our procedure for optimizing the cuts is as follows:  Begin with a
set of loose (pre-selection) cuts and choose cuts on
$\Delta\eta_{j_1j_2}$, $\Delta\eta_{ll}$, $\Delta\phi_{ll}$,
$M_{j_1j_2}$, $M_{ll}$, and the transverse mass,
$M_{T}$,\footnote{The transverse mass is defined as
in~\cite{pr_160_113004,pl_503_113,pr_61_093005,JHEP_9712_005}.}
that optimize $S/\sqrt{B}$, where S and B are the expected number
of signal and background events for $30\,$fb$^{-1}$ of luminosity,
respectively. We perform this optimization with a genetic
algorithm~\cite{GAlib}. We perform this procedure for several
masses and find a parametrization for the optimal cut as a
function of the Higgs mass.

The following event selection was chosen:
\begin{itemize}
\item[{\bf a.}] Topology cuts. Require two charged leptons
($e,\mu$) that pass the single or double charged lepton trigger in
ATLAS. Here, a veto on b-jets is applied (see
Section~\ref{sec:expsig} and~\cite{SN-ATLAS-2003-024}). \item[{\bf
b.}] Forward jet tagging with $P_{Tj_1},P_{Tj_2}>20\,\gev$ and
$\Delta\eta_{j_1j_2}^{min}<\Delta\eta_{j_1j_2}$ according to
\begin{equation}
\Delta\eta_{j_1j_2}^{min}={a\over(M_{H}-b)}+cM_{H}+d,
\end{equation}
where $a=2861$, $b=-327$, $c=9.6\times 10^{-3}$, and $d=-3.44$.
Leptons are required to be in between jets in pseudorapidity.
\item[{\bf c.}]  Tau
rejection~\cite{pr_160_113004,pl_503_113,pr_61_093005,JHEP_9712_005,SN-ATLAS-2003-024}.
\item[{\bf d.}]  Tagging jets invariant mass:
$520\,\gev<M_{j_1j_2}<3325\,\gev$ \item[{\bf e.}] Central jet veto
(see Section~\ref{sec:expsig} and~\cite{SN-ATLAS-2003-024}).
\item[{\bf f.}] Lepton angular cuts: We require
$\Delta\eta_{ll}<\Delta\eta_{ll}^{max}$ with
\begin{equation}
\Delta\eta_{ll}^{max}=a+bM_H+cM_H^2,
\end{equation}
where $a=6.25$, $b=-6.24\times 10^{-2}$, $c=1.99\times 10^{-4}$
for $M_H<200\,\gev$, and $a=3.88$, $b=-4.17\times 10^{-3}$, $c=0$
for $M_H>200\,\gev$. It is required that
$\Delta\phi_{ll}^{min}<\Delta\phi_{ll}<\Delta\phi_{ll}^{max}$ with
\begin{equation}
\Delta\phi_{ll}^{min}=a+bM_H,
\end{equation}
where $a=-2.20$, $b=7.54\times 10^{-3}$, and
\begin{equation}
\Delta\phi_{ll}^{max}=a+bM_H+cM_H^2+dM_H^3,
\end{equation}
where $a=-4.07$, $b=0.156559$, $c=-1.310956\times 10^{-3}$, and
$d=3.42011\times 10^{-6}$. As one would expect, the minimum cut is
only important at high Higgs masses, and the upper bound is only
relevant at low Higgs masses. It is
 required that $M_{ll}^{min}<M_{ll}<M_{ll}^{max}$ with
 \begin{equation}
M_{ll}^{min}=a (M_{H}-b)^{2}+c,
 \end{equation}
 where $a=-2.82\times 10^{-3}$, $b=464$, $c=129$, and
\begin{equation}
M_{ll}^{max}={{a (M_{H}-b)^{2}} \over {d + (M_{H}-b)^{2}}} +c,
\end{equation}
where $a=310$, $b=114$, $c=47.6$, and $d=13290$. In order to
further reduce the
 contribution from Drell-Yan, we
 require $85<M_{ll}<95\,\gev$ and $\sla{p_T}>30\,\gev$, if leptons are of same flavor.
\item[{\bf g.}] Transverse mass cuts. We require that
 $M_{T}^{min}<M_{T}<M_{T}^{max}$,  with
\begin{equation}
M_{T}^{min}=a+bM_{H},
\end{equation}
where $a=-17$ and $b=0.73$ and
\begin{equation}
M_{T}^{max}=a+bM_{H}+cM_{H}^{2}+dM_{H}^3,
\end{equation}
where $a=106$, $b=-0.83$, $c=9.46\times 10^{-3}$, and
$d=-9.49\times 10^{-6}$. We also require $m_T(ll\nu\nu)>30\,\gev$,
with
$m_T(ll\nu\nu)=\sqrt{2P^{ll}_T\sla{p_T}(1-\cos{\Delta\phi})}$,
where $P^{ll}_T$ is the $P_T$ of the di-lepton system and
$\Delta\phi$ corresponds to the angle between the di-lepton vector
and the $\sla{p_T}$ vector in the transverse plane.

\end{itemize}

\begin{table}[t]
\begin{center}
\begin{tabular}{c || c c c c}
\hline \hline
$M_{H}(\gev)$ & $e-\mu$ & $e-e$ & $\mu-\mu$ & Combined\\
\hline \hline
115     & 0.9   & 0.4   & 0.5   & 1.4\\
130     & 3.0   & 1.5   & 2.2   & 4.3\\
160     & 8.2   & 5.1   & 6.3   & 11.6\\
200     & 4.4   & 2.6   & 3.0   & 6.0\\
300     & 2.3   & 1.4   & 1.5   & 3.1\\
500     & 1.0   & 0.6   & 0.6   & 1.5\\
\hline \hline
\end{tabular}
\caption{Expected Poisson significance for the parameterized cuts
listed in Section~\ref{sec:hwwevsel} with 10\,fb$^{-1}$ of
integrated luminosity. A 10\% systematic uncertainty is applied to
all backgrounds when calculating the significance.
}\label{table:mcatnloSig}
\end{center}
\end{table}

\subsection{Results and Discovery Potential}
\label{sec:hwwres}

Table~\ref{tab:mcanloCuts} displays effective cross-sections for
signal and background after application of successive cuts
presented in Section~\ref{sec:hwwevsel}. Cross-sections are
presented for $M_H=160\,\gev$ in the $e-\mu$ channel. It is worth
noting that the central jet veto survival probability for
$t\overline{t}$ production is significantly lower than that
reported in~\cite{SN-ATLAS-2003-024}. However, this is compensated by a lower rejection due to requiring two tagging jets (see cut {\bf b} in the previous Section). As a result, the relative contribution to the background from $t\overline{t}$ production obtained here is similar to the one reported in~\cite{SN-ATLAS-2003-024}.
Table~\ref{table:mcatnloSig}
reports the expected Poisson significance for 10\,fb$^{-1}$ of
integrated luminosity. Simple event counting is used and a
$10\,\%$ systematic error on the background determination was
assumed. In order to incorporate the systematic errors we
incorporated~\cite{ATL-PHYS-2003-008,physics_03_12050} the
formalism developed in~\cite{nim_A320_331}. The implementation of
MC@NLO to simulate the $t\overline{t}$ background has not changed
the conclusions drawn in~\cite{SN-ATLAS-2003-024} for the $M_H$
considered there. The $H\rightarrow W^{(*)}W^{(*)}\rightarrow
l^+l^-\sla{p_T}$ mode has a strong potential in a wide rage of
Higgs masses. A significance of or greater than $5\,\sigma$ may be
achieved with 30\,fb$^{-1}$ of integrated luminosity for
$125<M_H<300\,\gev$.

\section{The $H\rightarrow\gamma\gamma$ Mode}
\label{sec:hgg}

\subsection{Generation of Background Processes}
\label{sec:ggbackground} The relevant background processes for
this mode are subdivided into two major groups. Firstly, the
production of two $\gamma$'s associated with two jets (real photon
production). Secondly, a sizeable contribution is expected from
events in which at least one jet is misidentified as a photon
(fake photon production). Despite the impressive jet rejection
rate after the application of $\gamma$ selection criteria expected
to be achieved by the ATLAS detector~\cite{LHCC99-14} ($\mayor
10^3$ for each jet), the contribution from fake photons will not
be negligible due to the large cross-sections of QCD processes at
the LHC.

LO ME based MC's were used to simulate $\gamma\gamma jj$ (both QCD
and EW diagrams), $\gamma jjj$ and $jjjj$ production. For this
purpose MadGraphII~\cite{pc_81_357,hep-ph_0208156} interfaced
with PYTHIA6.2 was used~\cite{Wisc_soft}. The factorization and
re-normalization scales were set to the $P_T$ of the lowest $P_T$
parton.

After the application of a number of basic cuts at the generator
level (see~\cite{ATL-PHYS-2003-036}) the QCD and EW $\gamma jjj$
diagrams correspond to 6.32 nb and 1.21 pb, respectively. Assuming
an effective jet rejection of the order of $10^{3}$, the starting
cross-section for the EW $\gamma jjj$ process would be about
$1\,$fb. This small cross-section will be severely reduced after
the application of further selection cuts (see
Section~\ref{sec:ggevsel}). In the analysis EW $\gamma jjj$ and
$jjjj$ diagrams were neglected.\footnote{More details of MC
generation for background processes are available
in~\cite{ATL-PHYS-2003-036}.}

\subsection{Event Selection}
\label{sec:ggevsel}

A number of pre-selection cuts are applied which are similar to
those used in the multivariate analysis of VBF $H\rightarrow
W^{(*)}W^{(*)} \rightarrow
l^{+}l^{-}\sla{p_{T}}$~\cite{ATL-PHYS-2003-007}:
\begin{itemize}
\item [{\bf a.}] $P_{T\gamma 1}, P_{T\gamma 2}>25\,\gev$. The
$\gamma$'s are required to fall in the central region of the
detector excluding the interface between the barrel and end-cap
calorimeters ($1.37<\left|\eta\right|<1.52$). The latter
requirement reduces the acceptance by about 10$\%$. \item [{\bf
b.}] Tagging jets with $P_{Tj_1}, P_{Tj_2}>20\,\gev$ and
$\Delta\eta_{j_1j_2}>3.5$. \item [{\bf c.}] The $\gamma$'s should
be in between the tagging jets in pseudorapidity. \item [{\bf d.}]
Invariant mass of the tagging jets, $M_{j_1j_2}>100\,\gev$. \item
[{\bf e.}] Central jet veto~\cite{SN-ATLAS-2003-024}. \item [{\bf
f.}] Invariant mass window:
$M_H-2\,\gev<M_{\gamma\gamma}<M_H+2\,\gev$.
\end{itemize}

The final event selection is obtained by means of maximizing the
Poisson significance for 30 fb$^{-1}$ of integrated luminosity for
$M_H=120\,\gev$. The maximization procedure is performed with the
help of the MINUIT program~\cite{cpc_10_343}. The following
variables are chosen: $P_{Tj_1}$, $P_{Tj_2}$,
$\Delta\eta_{j_1j_2}$, $\Delta\phi_{j_1j_2}$, $M_{j_1j_2}$,
$P_{T\gamma_1}$, $P_{T\gamma_2}$, and $\Delta\eta_{\gamma\gamma}$.

Due to the implementation of parton shower and hadronization
effects, the kinematics of the final state  will be somewhat
different from that  of the parton level analysis performed
in~\cite{Rainwaterthesis}. After the application of cut {\bf f} in
the pre-selection, the dominant background corresponds to QCD
$\gamma\gamma jj$ and the fake photon production, therefore, the
optimization process will be mainly determined by the kinematics
of these process together with that of the VBF signal.

\begin{table}[t]
\begin{center}
\begin{tabular}{l || c c c c}
\hline \hline
  Cut   & Pre-selection  & Parton Level & Optimization  \\
  \hline\hline
 {\bf a} &  $P_{T\gamma 1}, P_{T\gamma 2} >25\,\gev$  &     $P_{T\gamma 1}>50\,\gev$    &     $P_{T\gamma 1}>57\,\gev$   \\

& &  $P_{T\gamma 2}>25\,\gev$ &  $P_{T\gamma 2}>34\,\gev$ \\

& & & $\Delta\eta_{\gamma\gamma}<1.58$,
$\Delta\phi_{\gamma\gamma}<3$~rad \\ \hline

 {\bf b} &   $P_{Tj_1}, P_{Tj_2} >20\,\gev$  &  $P_{Tj_1}>40\,\gev$
 & $P_{Tj_1}>40\,\gev$ \\

 & & $P_{Tj_2}>20\,\gev$ & $P_{Tj_2}>29.5\,\gev$ \\

& $\Delta\eta_{j_1j_2}>3.5$ &  $\Delta\eta_{j_1j_2}>4.4$ &
$\Delta\eta_{j_1j_2}>3.9$  \\ \hline

{\bf d} &   $M_{j_1j_2}>100\,\gev$    &     -  &     $M_{j_1j_2}>610\,\gev$ \\
\hline\hline

  \end{tabular}
 \caption{Values of the cuts applied for different event selections (see Section~\ref{sec:ggevsel}).}
 \label{tab:cuts}
\end{center}
\end{table}

Initially, it has been verified that the inclusion of variables
additional to those considered in~\cite{Rainwaterthesis} improves
the signal significance. The addition of the photon related
variables $\Delta\eta_{\gamma\gamma}$ and
$\Delta\phi_{\gamma\gamma}$ improves the signal significance by
some $10-20\%$ depending on the Higgs mass. The implementation of
those two variables separately proves more efficient than the
combined $\Delta R_{\gamma\gamma}$. The inclusion of the hadronic
variable $\Delta\phi_{jj}$ does not noticeably increase the signal
significance.

Table~\ref{tab:cuts} shows the results of the optimization
together with the values of the cuts placed at the pre-selection
level and for the parton level analysis performed
in~\cite{Rainwaterthesis}. Due to the significant increase in the
background contribution compared to the parton level analysis, the
optimized event selection is significantly tighter, resulting into
reduced signal and background rates (see Section~\ref{sec:ggres}).
The increase of the background comes from the different choice of
the width of the mass window, the implementation of parton showers
for the estimation of the central jet veto probability and the
inclusion of fake photon events.

\subsection{Results and Discovery Potential}
\label{sec:ggres}

Here, we use  the event selection obtained in the optimization
procedure performed in Section~\ref{sec:ggevsel} (see
Table~\ref{tab:cuts}). The expected  signal and background
cross-sections corrected for acceptance and efficiency corrections
are shown in Table~\ref{tab:cross120} for a mass window around
$M_H=120\,\gev$ after the application of successive cuts.

The contribution from the fake photon background has been severely
reduced thanks to the inclusion of the photon angular variables.
The contribution from this background, however, is important. The
normalization of the fake photon background is subject to sizeable
systematic uncertainties. This is partly due to the uncertainty on
the determination of the fake photon rejection
rate~\cite{LHCC99-14}.

Figure~\ref{fig:mgg} shows the expected signal and background
effective cross-section (in fb) as a function of
$M_{\gamma\gamma}$ for $M_H=130\,\gev$. The dashed line shows the
total background contribution whereas the dotted line corresponds
to the real $\gamma\gamma$ background. The solid line displays the
expected contribution of signal plus background.
\begin{figure}[t]
{\centerline{\epsfig{figure=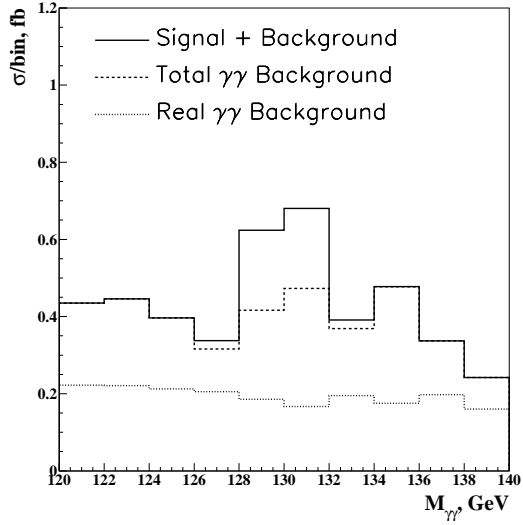,width=7.cm}}}
\vspace{-0.2cm} \caption[]{Expected signal and background
effective cross-section (in fb) as a function of
$M_{\gamma\gamma}$ for $M_H=130\,\gev$. The dashed line shows the
total background contribution whereas the dotted line corresponds
to the real $\gamma\gamma$ background. The solid line displays the
expected contribution of signal plus background.} \label{fig:mgg}
\end{figure}
In Table~\ref{tab:ggresults}, results are given in terms of $S$
and $B$, for 30 fb$^{-1}$ of integrated luminosity. The signal
significance was calculated with a Poissonian calculation. The
signal significance expected with this VBF mode alone reaches
2.2$\,\sigma$ for 30 fb$^{-1}$ of integrated luminosity.

The QCD $\gamma\gamma jj$ has been estimated with QCD
$\gamma\gamma jj$ ME based MC alone. The rate of additional (non
tagging) jets has been estimated with the help of the parton
shower. This approach yields a central jet veto survival
probability significantly smaller than that calculated
in~\cite{Rainwaterthesis}. Both effects go in the direction of
overestimating of the $\gamma\gamma jj$ background. Similar
discussion applies to the estimation of the fake photon background
performed here. This background estimation may be improved with
the implementation of a more realistic MC for the simulation of
the real photon background. This mode is considerably more sensitive to the understanding of fake photon rejection than the inclusive analysis~\cite{LHCC99-14}.

 \begin{table}[t]
 \begin{center}
 \begin{tabular}{l ||  c c c c c c}
 \hline \hline
 Cut & VBF H & g-g Fusion H & QCD $\gamma\gamma jj$ &      EW $\gamma\gamma jj$ & $\gamma jjj$ & $jjjj$ \\
\hline\hline
 {\bf a} &       2.25 &       5.45 &     246.90 &       7.97 &     172.60 &     691.06  \\
 {\bf b} &       0.73 &       0.08 &      31.83 &       4.39 &      28.30 &      35.22  \\
 {\bf c} &       0.70 &       0.07 &      16.81 &       4.20 &      21.76 &      30.06  \\
 {\bf d} &       0.57 &       0.04 &       7.43 &       3.69 &      12.77 &      16.99  \\
 {\bf e} &       0.42 &       0.02 &       5.41 &       2.50 &       8.52 &       8.49  \\
 {\bf f} &       0.38 &       0.02 &       0.28 &       0.14 &       0.22 &       0.25  \\
 \hline\hline
 \end{tabular}
 \caption{Expected signal and background cross-sections (in fb) corrected for acceptance and efficiency corrections after the application of successive cuts (see Section~\ref{sec:ggevsel}). Here $M_H=120\,\gev$.}
 \label{tab:cross120}
 \end{center}
 \end{table}

 \begin{table}[t]
 \begin{center}
\begin{tabular}{c || c c c c}
 \hline\hline
 $M_H$ & $S$ & $B$ & $S/B$ & $\sigma_P$     \\
\hline\hline
  110 &      10.05 &      30.69 &       0.33 &     1.56  \\
  120 &      12.06 &      26.54 &       0.45 &     2.02  \\
  130 &      12.52 &      23.97 &       0.52 &     2.19  \\
  140 &      10.91 &      22.90 &       0.48 &     1.94  \\
  150 &       7.69 &      20.15 &       0.38 &     1.42  \\
  160 &       2.89 &      17.21 &       0.17 &     0.44  \\
 \hline \hline
 \end{tabular}
 \caption{Expected number of signal and background events, $S/B$ and the corresponding signal significance for 30~fb$^{-1}$ of integrated luminosity (see Section~\ref{sec:ggres}).}
 \label{tab:ggresults}
 \end{center}
 \end{table}

\section{The $H\rightarrow ZZ\rightarrow l^+l^-q\overline{q}$ Mode}
\label{sec:hzz}

\subsection{Generation of Background Processes}
\label{sec:zzback}

Cross-section for the QCD $Z+4j, Z\rightarrow l^+l^-, l=e,\mu$
process were calculated with two independent packages:
ALPGEN~\cite{hep-ph_02_06293} and
MadGraphII~\cite{pc_81_357,hep-ph_0208156}. Both calculations
include the $Z/\gamma^{\star}$ interference effects. The following
cuts at the generator level were used for the cross-section
calculation for the nominal event generation:
\begin{itemize}
\item QCD parton's transverse momentum, $P_T>20\,\gev$,
pseudorapidity, $\left|\eta\right|<5$. Separation between QCD
partons, $\Delta R>0.5$. \item Minimal transverse momentum cuts on
leptons, $P_T>3\,\gev$ with $\left|\eta\right|<3$. The angle
separation between leptons and leptons and jets were set to
$\Delta R>0.2$
\end{itemize}

The Born level cross-section of QCD $Z+4j$ production is subject
to large uncertainties. Some properties of jets in association
with  $W$ and $Z$ bosons have been studied and have been compared
with QCD predictions at the
Tevatron~\cite{prl_77_448,prl_79_4760}. The measured
cross-sections of $W/Z+n~jets$ where $n=1,2,3,4$ lie in between
the LO predictions calculated using the re-normalization and
factorization scales equal to the average transverse momentum of
the partons, $\langle P_T\rangle$,  and the transverse energy of
the weak boson, $E_T^{WB}$, respectively. The LO prediction
calculated with the first choice of scale systematically
undershoots the measured cross-section. At the LHC $\langle
P_T\rangle>100\,\gev$, due to the large phase space. Thus, the
scale was set to the mass of the weak boson.

After the application of the cuts at the generator level and the
choice of scales mentioned above both ALPGEN and MadGraphII yield
$47.5\,$pb. 8.5 million un-weighted events were generated with
MadGraphII. The output from MadGraphII was interfaced to the
HERWIG6.5 package~\cite{Wisc_soft}. In order to avoid severe
double counting in the generation of hadronic jets, the scale of
the parton shower evolution was set to the $P_T$ of the lowest
transverse momentum parton in the event.

The cross-section for  $Z+4j, Z\rightarrow l^+l^-, l=e,\mu$
production with one EW boson in the internal lines was evaluated
with MadGraphII. These diagrams include QCD $ZZjj$ and $ZW^\pm
jj$. A cross-section of $1.6\,$pb was obtained after cuts at
generator level and by applying the same choice of scales as for
the QCD $Z+4j$ case. The impact of these diagrams is small, hence,
they were not included in the final results reported in
Section~\ref{sec:hzzresults}. Diagrams with two EW bosons in the
internal lines were not considered, as they are expected to be
negligible.

A sample of events for $t\overline{t}$ production was used. These
events were generated with the MC@NLO package (see
Section~\ref{sec:wwbackround}).

\subsection{Event Selection} \label{sec:hzzevsel}

The event selection presented in this Section is obtained by
maximizing the signal significance for a Higgs for $M_H=300\,\gev$
with $30\,$fb$^{-1}$ of integrated luminosity.

\begin{figure}[t]
{\centerline{\epsfig{figure=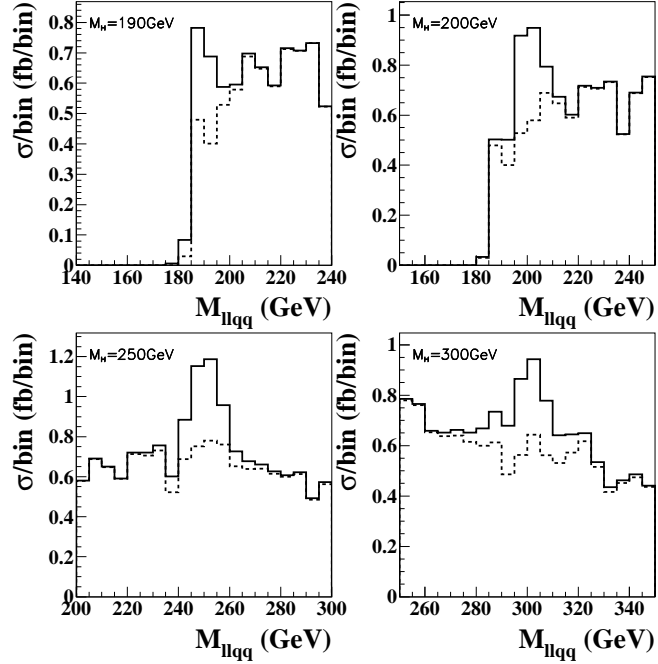,width=9.cm}}}
\vspace{-0.2cm} \caption[]{Invariant mass of the Higgs candidates
after the application of kinematic fits. The solid lines
correspond to the sum of the signal (VBF $H\rightarrow
ZZ\rightarrow l^+l^-q\overline{q}$) and the main background (QCD
$Z+4j, Z\rightarrow l^+l^-, l=e,\mu$). The dashed lines show the
contribution of the main background alone. Here $M_H=190, 200,
250, 300\,\gev$.} \label{fig:massplots_af_1}
\end{figure}

A number of basic features common to VBF modes remain.  A feature
specific to the mode under study is the additional ambiguity in
the definition of tagging jets introduced by the presence of
relatively hard jets produced from the decay of the $Z$'s.  A
search for two jets with an invariant mass close to $Z$ mass,
$M_Z$, is performed. After reconstructing the $Z$ decaying
hadronicaly, the event looks like a ``typical'' VBF candidate.

The following event selection was chosen:
\begin{itemize}
\item[{\bf a}.] Two isolated, oppositely charged, of equal flavor
leptons in the central detector region, $\left|\eta\right|<2.5$.
\item[{\bf b}.] The event is required to pass the single or double
lepton trigger in ATLAS. \item[{\bf c}.] Two hadronic jets ($j_3,
j_4$) with transverse momentum, $P_T>30\,\gev$  with $M_{j_3j_4}$
close to $M_Z$ were required in the fiducial region of the
calorimeter, $\left|\eta\right|<4.9$. The relative invariant mass
resolution of two jets is expected to be approximately $10\,\%$.
The following mass window was chosen: $75<M_{j_3j_4}<101\,\gev$.
These jest were ``masked out'' from the list of jets. \item[{\bf
d}.] Tagging jets with $P_{Tj_1}>40\,\gev$, $P_{Tj_2}>30\,\gev$
and $\Delta\eta_{j_1j_2}>4.4$. \item[{\bf e}.] Both leptons were
required to lie in between the tagging jets in pseudorapidity.
\item[{\bf f}.] Leptonic cuts. It was required that
$M_Z-10<M_{ll}<M_Z+10\,\gev$. This cut is expected to suppress
di-lepton final states with $W^+W^-\rightarrow ll\nu\nu$. It is
particularly important to suppress the contribution from
$t\overline{t}$ production associated with jets. No b-tagging
rejection algorithms were applied in this analysis due to the
large branching ratio of $Z$ decaying into heavy quarks.
\item[{\bf g}.] The invariant mass of the tagging jets was
required to be greater than $900\,\gev$. \item[{\bf h}.] Central
jet veto. Extra jets with $P_T>20\,\gev$ are looked for in the
central region of the detector ($\left|\eta\right|<3.2$). However,
high $P_T$ quarks from the decay of one of the $Z$'s are expected
to radiate hard gluons with a high probability, thus, faking
hadronic jets produced prior to the decay. If $\Delta R$ between
the extra jet and the jets of the Higgs candidate is larger than
one unit, the event is vetoed. \item[{\bf i}.] In order to further
reduce the contribution from events with $W^+W^-\rightarrow
ll\nu\nu$, it is required that $\sla{p_T}<30\,\gev$.
\end{itemize}

The $M_{llj_3j_4}$ spectrum could be distorted due to the
ambiguity in defining tagging jets. The distortion of the
$M_{llj_3j_4}$ spectrum, however, is not sizeable.
Figure~\ref{fig:massplots_af_1} displays the $M_{llj_3j_4}$
spectra for signal and background after the application of the
event selection presented in this Section. A Higgs mass resolution
of approximately $2.5\%$ is obtained for
$2M_Z<M_H<300\,\gev$~\cite{ATL-COM-PHYS-2003-035}.

\subsection{Results and Discovery Potential}
\label{sec:hzzresults}

Table~\ref{tab:hzzcrosec} shows the expected signal effective
cross-sections (in fb) for a Higgs mass of $M_H=300\,\gev$.
Table~\ref{tab:hzzcrosec} also displays the effective
cross-sections for the major background processes. Cross-sections
are given after successive cuts (see Section~\ref{sec:hzzevsel}).
The background is largely dominated by the QCD $Z+4j, Z\rightarrow
l^+l^-, l=e,\mu$ production. Diagrams with one or two EW boson in
the internal lines were neglected. The contribution from
$t\overline{t}$ is small and it is also neglected in the final
results.

\begin{table}[t]
 \begin{center}
 \begin{tabular}{l || c c c c c c c c c}
 \hline\hline
 Process & {\bf a} & {\bf b} & {\bf c} & {\bf d} & {\bf e} & {\bf f} & {\bf g} &{\bf h}& {\bf i}\\
\hline\hline
 VBF ($ M_H=300\,\gev$) &      31.69 &      31.50 &      12.63 &       3.39 &       3.26 &       2.93 &       2.24 &       1.72 &       1.66\\
QCD $Z+4j$ &   25930 &      25902 &  10345   &   277  &  205   &    205 &   116 &  36.6 &  34.6  \\
  $t\overline{t}$ &  14793 &   14268 &   4233 &   135 &   106 &    10.5  &   6.4   &   2.3 &  0.3  \\
 \hline\hline
 \end{tabular}
\caption{Expected effective cross-sections (in fb) for
$H\rightarrow ZZ\rightarrow llq\overline{q}$ produced via VBF
$(M_H=300\,\gev$) and the main background processes.
Cross-sections are given after successive cuts presented in
Section~\ref{sec:hzzevsel}.}
 \label{tab:hzzcrosec}
 \end{center}
 \end{table}

Table~\ref{tab:results_af} reports results in terms of $S$, $B$,
$S/B$ and signal significance, $\sigma_L$, with 30~fb$^{-1}$ of
integrated luminosity for different values of $M_H$. The effective
signal and background cross-sections are evaluated in a
$4\,\sigma_M$ (where $\sigma_M$ is the mass resolution) wide mass
window. The signal significance was calculated with a likelihood
ratio technique using the invariant mass of the Higgs candidate as
a discriminant
variable~\cite{ATL-PHYS-2003-008,physics_03_12050}. A signal
significance of $3.75\,\sigma$ may be achieved for $M_H=300\,\gev$
with 30\,fb$^{-1}$ of integrated luminosity. It should be noted
that the cross-sections for the main background reported here are
subject to large theoretical uncertainty. Fortunately, the
background may be determined from side bands for Higgs searches
with $M_H>200\,\gev$.

 \begin{table}[t]
 \begin{center}
 \begin{tabular}{ c || c c c c}
 \hline \hline
$M_H (\gev)$ & $S$ & $B$ & $S/B$ & $\sigma_L$ \\
\hline\hline
 190 & 18.9 & 31.2 & 0.61 & 3.47 \\
 200 & 27.3 & 52.8 & 0.52 & 3.76 \\
 300 & 39.3 & 116.1 & 0.34 & 3.75 \\
 500 & 20.1 & 124.2 & 0.16 & 1.98 \\
\hline\hline
 \end{tabular}
 \caption{Expected number of signal and background events,  ratio of signal to
background and signal significance (in $\sigma$) for a SM Higgs
produced via VBF using the decay mode $H\rightarrow ZZ\rightarrow
l^+l^-q\overline{q}$ with 30~fb$^{-1}$ of integrated luminosity
for different values of $M_H$. The effective signal and background
cross-sections are evaluated in a $4\,\sigma_M$ (where $\sigma_M$
is the mass resolution) wide mass window. The signal significance,
$\sigma_L$, was calculated with a likelihood ratio technique using
the invariant mass of the Higgs candidate as a discriminant
variable.}
 \label{tab:results_af}
 \end{center}
 \end{table}

\section{Multivariate Analysis}
\label{sec:NN}

Results reported in~\cite{SN-ATLAS-2003-024} and the present paper
were based on classical cut analyses. Multivariate techniques have
been extensively used in physics analyses, for instance, in LEP
experiments. Neural Networks (NN) are the most commonly used tools
in multivariate analyses. NN training has been performed on the
$H\rightarrow W^{(*)}W^{(*)}\rightarrow
l^+l^-\sla{p_T}$~\cite{ATL-PHYS-2003-007} and $H\rightarrow
\tau^+\tau^-\rightarrow l^+l^-\sla{p_T}$~\cite{NNVBFtautaull}
modes. NN training was performed with a relatively small number of
variables. It was required that these variables are infra-red safe
and their correlations do not depend strongly on detector effects:
$\Delta\eta_{j_1j_2}$, $\Delta\phi_{j_1j_2}$, $M_{j_1j_2}$,
$\Delta\eta_{ll}$, $\Delta\phi_{ll}$, $M_{ll}$, and $M_{T}$ (or
the invariant mass of the $\tau^+\tau^-$ system in the case of the
$H\rightarrow \tau^+\tau^-\rightarrow l^+l^-\sla{p_T}$ mode). The
signal significance was calculated with a likelihood ratio
technique using the NN output as a discriminant variable. An
enhancement of approximately $30-50\,\%$ of the signal
significance with respect to the classical cut analysis was
obtained for both modes under consideration.

\section{Conclusions}
\label{sec:conclusions}

The discovery potential for the SM Higgs boson produced with VBF
in the range $115<M_H<500\,\gev$ has been reported. An updated
study at hadron level followed by a fast detector simulation of
the $H\rightarrow W^{(*)}W^{(*)}\rightarrow l^+l^-\sla{p_T}$ mode
has been presented: the main background, $t\overline{t}$
associated with jets, has been modelled with the MC@NLO program
and the Higgs mass range has been extended to $500\,\gev$. This
mode has a strong potential: a signal significance of more than
$5\,\sigma$ may be achieved with 30\,fb$^{-1}$ of integrated
luminosity for $125<M_H<300\,\gev$. The discovery potential of the
$H\rightarrow\gamma\gamma$ and $H\rightarrow ZZ\rightarrow
l^+l^-q\overline{q}$ modes have also been reported with analyses
at hadron level followed by a fast detector simulation.

The discovery potential of the modes presented in this work was
combined with results reported in past studies performed for the
ATLAS
detector. Results from recent studies~\cite{ATL-PHYS-2003-001,ATL-PHYS-2003-024,ATL-PHYS-2003-025}, which were not used in~\cite{SN-ATLAS-2003-024}, were added here.
Likelihood ratio techniques have been used to perform the
combination~\cite{ATL-PHYS-2003-008,physics_03_12050}. In order
to incorporate systematic errors, the formalism developed
in~\cite{nim_A320_331} was implemented. A 10\,\%
systematic error on the background estimation has been assumed for
modes related to VBF~\cite{SN-ATLAS-2003-024}. Figure~\ref{fig:discoveryPlot} displays the
overall discovery potential of the ATLAS detector with
10\,fb$^{-1}$ of integrated luminosity. Results from NN
based analyses and discriminating variables have not been included
in the combination. The present study confirms the results
reported
in~\cite{pr_160_113004,pl_503_113,pr_61_093005,JHEP_9712_005,SN-ATLAS-2003-024},
that the VBF mechanism yields a strong discovery potential at the
LHC in a wide range of the Higgs boson mass.

\begin{figure}[t]
\begin{center}
\epsfig{file=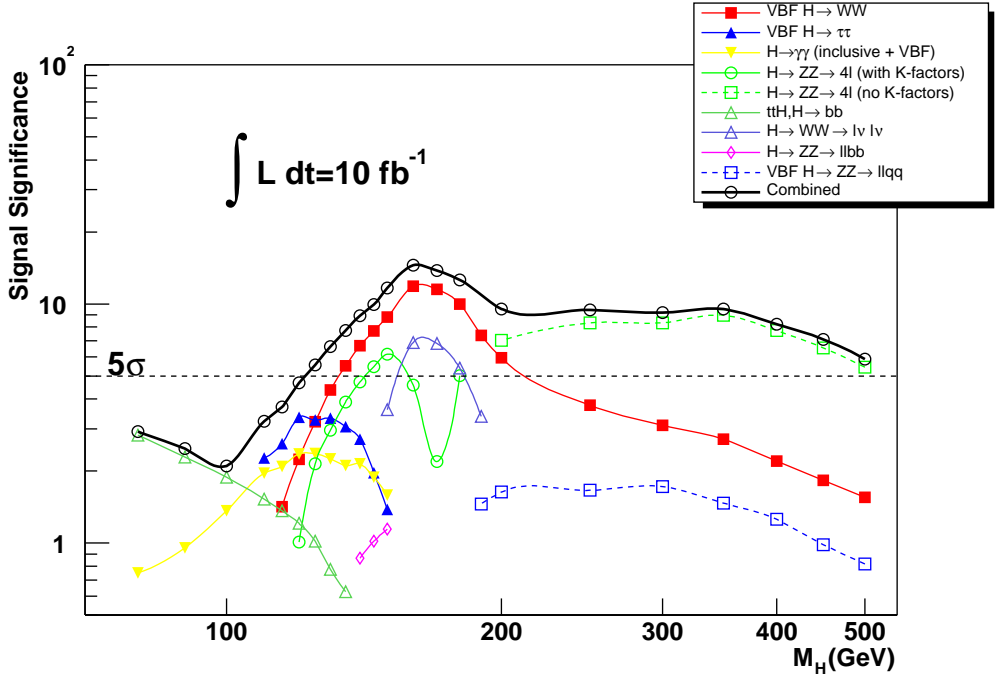, width=13.5cm}
\vspace{-0.2cm} \caption{Expected significance for ATLAS as a
function of Higgs mass for 10\,fb$^{-1}$ of integrated
luminosity.} \label{fig:discoveryPlot}
\end{center}
\end{figure}

\section{Acknowledgements}

We are particularly indebted to F.~Cerutti, S.~Frixione, K.~Jakobs,  T.~Plehn,
D.~Rainwater, and D.~Zeppenfeld. All of us would like to thank the
organizers of the {\it Les Houches Workshop} for the fruitful
workshop.

}

\pagebreak

\bibliographystyle{zeusstylem}
\bibliography{vbf,mycites}

\begin{mcbibliography}{10}

\bibitem{np_22_579}
S.~L.~Glashow,
\newblock  Nucl. Phys. {\bf B22}  (1961)~ 579\relax
\relax
\bibitem{prl_19_1264}
S.~Weinberg,
\newblock  Phys. Rev. Lett. {\bf 19}  (1967)~ 1264\relax
\relax
\bibitem{sal_1968_bis}
A.~Salam,
\newblock  Proceedings to the Eigth Nobel Symposium, May 1968, ed: N.~Svartholm
  (Wiley, 1968) 357\relax
\relax
\bibitem{pr_2_1285}
S.L.~Glashow, J.~Iliopoulos and L.~Maiani,
\newblock  Phys. Rev. {\bf D2}  (1970)~ 1285\relax
\relax
\bibitem{pl_12_132}
P.W.~Higgs,
\newblock  Phys. Lett. {\bf 12}  (1964)~ 132\relax
\relax
\bibitem{prl_13_508}
P.W.~Higgs,
\newblock  Phys. Rev. Lett. {\bf 13}  (1964)~ 508\relax
\relax
\bibitem{pr_145_1156}
P.W.~Higgs,
\newblock  Phys. Rev. {\bf 145}  (1966)~ 1156\relax
\relax
\bibitem{prl_13_321}
F.~Englert, R.~Brout,
\newblock  Phys. Rev. Lett. {\bf 13}  (1964)~ 321\relax
\relax
\bibitem{prl_13_585}
G.S.~Guralnik, C.R.~Hagen and T.W.B.~Kibble,
\newblock  Phys. Rev. Lett. {\bf 13}  (1964)~ 585\relax
\relax
\bibitem{pr_155_1554}
T.W.B.~Kibble,
\newblock  Phys. Rev. {\bf 155}  (1967)~ 1554\relax
\relax
\bibitem{LHCC99-14}
ATLAS Collaboration,
\newblock  Detector and Physics Performance Technical Design Report,
\newblock  CERN-LHCC/99-14 (1999)\relax
\relax
\bibitem{prl_40_11_692}
H.M.~Georgi, S.L.~Glashow, M.E.~Machacek and D.V.~Nanopoulos,
\newblock  Phys. Rev. Lett. {\bf 40}  (1978)~ 11\relax
\relax
\bibitem{pl_136_196}
R.~Cahn and S.~Dawson,
\newblock  Phys. Lett. {\bf B136}  (1984)~ 196\relax
\relax
\bibitem{pl_148_367}
G.~Kane, W.~Repko and W.~Rolnick,
\newblock  Phys. Lett. {\bf B148}  (1984)~ 367\relax
\relax
\bibitem{pr_160_113004}
D.L.~Rainwater and D.~Zeppenfeld,
\newblock  Phys. Rev. {\bf D60}  (1999)~ 113004\relax
\relax
\bibitem{pl_503_113}
N.~Kauer, T.~Plehn, D.L.~Rainwater and D.~Zeppenfeld,
\newblock  Phys. Lett. {\bf B503}  (2001)~ 113\relax
\relax
\bibitem{pr_61_093005}
T.~Plehn, D.L.~Rainwater and D.~Zeppenfeld,
\newblock  Phys. Rev. {\bf D61}  (2000)~ 093005\relax
\relax
\bibitem{JHEP_9712_005}
D.L.~Rainwater and D.~Zeppenfeld,
\newblock  JHEP {\bf 9712} (1997) 005\relax
\relax
\bibitem{SN-ATLAS-2003-024}
S.~Asai {\it et al.},
\newblock  Search for the Standard Model Higgs Boson in ATLAS using Vector
  Boson Fusion,
\newblock  ATLAS Scientific Note SN-ATLAS-2003-024 (2003), submitted to EPJ,
  hep-ph/0402254\relax
\relax
\bibitem{cpc_82_74}
T.~Sj\"ostrand,
\newblock  Comp. Phys. Comm. {\bf 82} (1994) 74\relax
\relax
\bibitem{cpc_135_238}
T.~Sj\"ostrand {\it et al.},
\newblock  Comp. Phys. Comm. {\bf 135} (2000) 238\relax
\relax
\bibitem{VV2H}
M.~Spira,
\newblock  VV2H Programe,
\newblock  home.cern.ch/m/mspira/www.proglist.html\relax
\relax
\bibitem{epj_12_375}
H.L.~Lai {\it et al.},
\newblock  Eur. Phys. J. {\bf C12}  (2000)~ 375\relax
\relax
\bibitem{cpc_108_56}
A.~Djouadi, J.~Kalinowski and M.~Spira,
\newblock  HDECAY: a Program for Higgs Boson Decays in the Standard Model and
  its Supersymmetric Extention,
\newblock  Comp. Phys. Comm. {\bf 108} (1998) 56\relax
\relax
\bibitem{ATLFAST}
E.~Richter-Was, D.~Froidevaux and L.~Poggioli,
\newblock  ATLFAST2.0 a Fast Simulation Package for ATLAS,
\newblock  ATLAS Note ATL-PHYS-98-131 (1998)\relax
\relax
\bibitem{ButtarHarperJakobs}
C.~Buttar, K.~Jakobs and R.~Harper,
\newblock  Weak boson fusion H-WW(*)-l+l-Pt-miss as a search mode for an
  intermediate mass SM Higgs boson at ATLAS,
\newblock  ATLAS Note ATL-PHYS-2002-033 (2002)\relax
\relax
\bibitem{JHEP_0206_029}
S.~Frixione and B.R.~Webber,
\newblock  JHEP {\bf 0206} (2002) 029\relax
\relax
\bibitem{JHEP_0308_007}
S.~Frixione and B.R.~Webber,
\newblock  JHEP {\bf 0308} (2003) 007\relax
\relax
\bibitem{JHEP_0101_010}
G.~Corcella {\it et al.},
\newblock  JHEP {\bf 0101} (2001) 010\relax
\relax
\bibitem{HERWIG6.5}
G.~Corcella {\it et al.},
\newblock  HERWIG 6.5: an Event Generator for Hadron Emission Reactions with
  Interfering Gluons,
\newblock  hep-ph/0011363\relax
\relax
\bibitem{pc_81_357}
T.~Stelzer and W.F.~Long,
\newblock  Phys. Comm. {\bf 81} (1994) 357\relax
\relax
\bibitem{hep-ph_0208156}
F.~Maltoni and T.~Stelzer,
\newblock  MadEvent: Automatic Event Generation with MadGraph,
\newblock  hep-ph/0208156 (2002)\relax
\relax
\bibitem{Wisc_soft}
Wisconsin Higgs Project,
\newblock  Sotfware web page,
\newblock  \\ http://www-wisconsin.cern.ch/physics/software.html\relax
\relax
\bibitem{ATL-COM-PHYS-2003-043}
Y.Q.~Fang, B.~Mellado, W.~Quayle, S.~Paganis and Sau Lan Wu,
\newblock  A Study of the $t\overline{t}+jets$ background at LHC,
\newblock  ATLAS internal Communication ATL-COM-PHYS-2003-043 (2003)\relax
\relax
\bibitem{zeppenfeld_1}
D.~Zeppenfeld,
\newblock  private communication\relax
\relax
\bibitem{Mazini}
R.~Mazini,
\newblock  private communication\relax
\relax
\bibitem{MadCUP}
D.~Zeppendeld {\it et al.},
\newblock  The Madison Collection of User Processes,
\newblock  http://pheno.physics.wisc.edu/Software/MadCUP/\relax
\relax
\bibitem{GAlib}
M.~Wall,
\newblock  GAlib: A C++ Library of Genetic Algorithm Components,
\newblock  http://lancet.mit.edu/ga/\relax
\relax
\bibitem{ATL-PHYS-2003-008}
K.~Cranmer, B.~Mellado, W.~Quayle and Sau Lan Wu,
\newblock  Confidence Level Calculations in the Search for Higgs Bosons Decay
  $H\rightarrow W^+W^- \rightarrow l^{+}l^{-}\sla{p_{T}}$ Using Vector Boson
  Fusion,
\newblock  ATLAS Note ATL-PHYS-2003-008 (2003)\relax
\relax
\bibitem{physics_03_12050}
K.~Cranmer, B.~Mellado, W.~Quayle and Sau Lan Wu,
\newblock  Challenges of Moving the LEP Higgs Statistics to the LHC,
\newblock  physics/0312050 (2003)\relax
\relax
\bibitem{nim_A320_331}
R.D.~Cousins and V.L. Highland,
\newblock  Nucl. Instrum. Methods {\bf A320}  (1992)~ 331\relax
\relax
\bibitem{ATL-PHYS-2003-036}
K.~Cranmer, B.~Mellado, W.~Quayle and Sau Lan Wu,
\newblock  Search for Higgs Boson Decay $H\rightarrow\gamma\gamma$ Using Vector
  Boson Fusion,
\newblock  ATLAS Note ATL-PHYS-2003-036 (2003), hep-ph/0401088\relax
\relax
\bibitem{ATL-PHYS-2003-007}
K.~Cranmer, B.~Mellado, W.~Quayle and Sau Lan Wu,
\newblock  Neural Network Based Search for Higgs Boson Produced via VBF with $H
  \rightarrow W^+W^- \rightarrow l^+ l^- \sla{p_{T}}$ for $115<M_H<130~\gev$,
\newblock  ATLAS Note ATL-PHYS-2003-007 (2003)\relax
\relax
\bibitem{cpc_10_343}
F.~James and M.~Roos,
\newblock  Comp. Phys. Comm. {\bf 10} (1975) 343\relax
\relax
\bibitem{Rainwaterthesis}
D.L.~Rainwater,
\newblock  Intermediate-Mass Higgs Searches in Weak Boson Fusion,
\newblock  Ph.D. thesis, University of Wisconsin - Madison, 1999\relax
\relax
\bibitem{hep-ph_02_06293}
M.L.~Mangano {\it et al.},
\newblock  ALPGEN, a Generator for Hard Multiparton Processes in Hadronic
  Collisions,
\newblock  hep-ph/0206293 (2002)\relax
\relax
\bibitem{prl_77_448}
CDF Collaboration, F.~Abe {\it et al.},
\newblock  Phys. Rev. Lett. {\bf 77}  (1996)~ 448\relax
\relax
\bibitem{prl_79_4760}
CDF Collaboration, F.~Abe {\it et al.},
\newblock  Phys. Rev. Lett. {\bf 79}  (1997)~ 4760\relax
\relax
\bibitem{ATL-COM-PHYS-2003-035}
K.~Cranmer, B.~Mellado, W.~Quayle and Sau Lan Wu,
\newblock  Search for Higgs Boson Decay $H\rightarrow ZZ\rightarrow l^+l^- qq,
  l=e,\mu$ with $2\cdot M_Z<M_H<500\,\gev$ Using Vector Boson Fusion,
\newblock  ATLAS internal Communication ATL-COM-PHYS-2003-035 (2003)\relax
\relax
\bibitem{NNVBFtautaull}
K.~Cranmer, B.~Mellado, W.~Quayle and Sau Lan Wu,
\newblock  Neural Network Based Search for Higgs Boson Produced via VBF with $H
  \rightarrow \tau^+\tau^- \rightarrow l^+ l^- \sla{p_{T}}$,
\newblock  ATLAS Note, in preparation\relax
\relax
\bibitem{ATL-PHYS-2003-001}
G.~Martinez, E.~Gross, G.~Mikenberg and L.~Zivkovic,
\newblock  Prospects for Light Higgs Observation in the $H^0\rightarrow
  Z^0Z^{0\star}\rightarrow b\overline{b}ee(\mu\mu)$ Channel at the LHC,
\newblock  ATLAS Note ATL-PHYS-2003-001 (2003)\relax
\relax
\bibitem{ATL-PHYS-2003-024}
J.~Cammin and M.~Schumacher,
\newblock  The ATLAS Discovery Potential for the Channel ttH, H to bb,
\newblock  ATLAS Note ATL-PHYS-2003-024 (2003)\relax
\relax
\bibitem{ATL-PHYS-2003-025}
K.~Cranmer, B.~Mellado, W.~Quayle and Sau Lan Wu,
\newblock  Application of K Factors in the $H\rightarrow ZZ^{\star}\rightarrow
  4l$ Analysis at the LHC,
\newblock  ATLAS Note ATL-PHYS-2003-025 (2003),
\newblock  hep-ph/0307242\relax
\relax
\end{mcbibliography}

\end{document}